# When supporting electrolyte matters – tuning capacitive response of graphene oxide *via* electrochemical reduction in alkali and alkaline earth metal chlorides

Dalibor Karačić[1], Selma Korać[2], Igor A. Pašti[1]*, Sanjin J. Gutić[2]

[1]*University of Belgrade – Faculty of Physical Chemistry, Belgrade, Serbia*

[2]*University of Sarajevo, Faculty of Science, Department of Chemistry, Sarajevo, Bosnia and Herzegovina*

**Abstract**

The ability to tune charge storage properties of graphene oxide (GO) is of utmost importance for energy conversion applications. Here we show that electrochemical reduction of GO is highly sensitive to the cations present in the solution. GO is reduced at lower potential in alkali metal chloride solutions than in alkaline earth metal chlorides. During the reduction, capacitance of GO increases from 10 to 70 times. Maximum capacitances of reduced GO are between 65 and 130 F g$^{-1}$, depending on the electrolyte and the presence of conductive additive. We propose that different interactions of cations with oxygen functional groups of GO during the reduction are responsible for the observed effect.

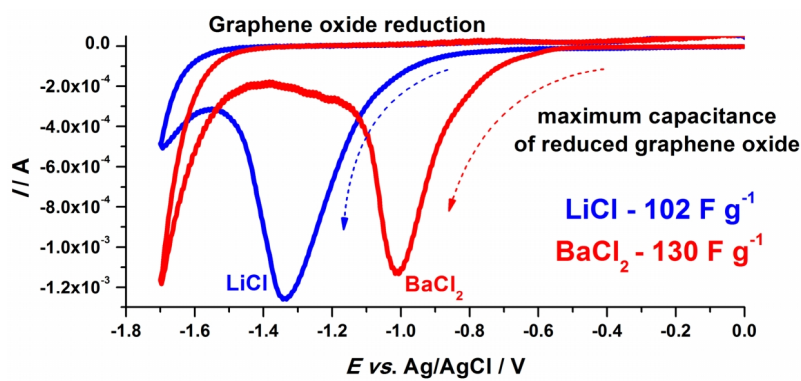

---

* **corresponding author:** Dr. Igor A. Pašti, e-mail: igor@ffh.bg.ac.rs

# 1. Introduction

Recent years witnessed rapid increase in knowledge on graphene and graphene-based materials, as well as the emergence of new technologies for a cheap and scalable production of different graphene-based nanostructures [1,2]. This is driven not only by the scientific curiosity, but also by the potential applications of these materials in almost all the areas of modern science and technology. Unique properties of graphene-based materials important for energy storage and conversion are already well acknowledged and used in different systems [3-5]. Among different methods for graphene production, wet chemical methods enable cheap production of large quantities of the „low-quality" graphene [6]. However, from the electrochemical point of view, this „low-quality" graphene is the most interesting as the functional groups and vacancies on graphene basal plane act as the active sites for different processes important for the application in electrochemical systems. In terms of charge storage application of graphene based materials, it is accepted that oxygen and nitrogen functional groups [5,7-10] contribute to capacitance through pseudocapacitive processes. Some recent works discuss the attempts for a precise modification of oxygen functionalities on graphene and graphene oxide (GO) basal plane, in order to finely tune its physical, chemical and electrochemical properties [9,11-14]. Among different proposed methods, electrochemical reduction of GO seems to be competitive approach for such modifications, due to a simple control of the reduction process through the control of the potential and reduction time, scalability and the avoidance of toxic reducing agents [15-18]. This places electrochemical reduction into the „green zone". Number of different experimental conditions and their effects on the reduction process and the properties of final reduced GO (rGO) can be found in literature [15-21].

Recently, we have shown general trends in tuning capacitive properties of GO [19]. As GO is reduced it gets de-oxygenated, while its conductivity increases. Capacitance is maximized when the concentration of the oxygen functional groups is properly balanced with the conductivity. The observed trends were shown for a number of different GO materials. While reduction of GO is pH dependent [20,21], the question is whether the evolution of GO to rGO during electrochemical reduction can be affected by other factors as well. In this communication, the effects of the alkali metal and alkaline earth metal cations on the electrochemical reduction of GO film are investigated. We show that the reduction of GO is affected by the composition of the inert electrolyte which also reflects on the capacitive properties of such obtained rGO.

## 2. Experimental part

Aqueous graphene oxide suspension (standard solution, 4 mg ml$^{-1}$; Graphenea, Spain [22]) was diluted to obtain 1 mg ml$^{-1}$ in 6:4 water/ethanol mixture. Also, we prepared analogous GO dispersion containing 0.1 mg ml$^{-1}$ of Vulcan XC-72R as conductive additive. For each experiment, after intensive sonication, 10 µl of GO suspension was drop-casted onto the glassy carbon disc (0.196 cm$^2$) and dried under the vacuum at room temperature. Standard three-electrode cell, with Pt foil (1 cm$^2$) as a counter and Ag/AgCl (saturated KCl) as a reference electrode, was used. All potentials are given versus this reference. PAR 263A potentiostat/galvanostat, controlled by PowerSUITE interface, was used.

Electrochemical measurements were performed in aqueous solutions of alkali (Li, Na, K, Rb, and Cs) and alkaline earth metal (Mg, Ca, Sr, Ba) chlorides (0.1 mol dm$^{-3}$), with pH adjusted to 5.5±0.1. After the GO electrode was prepared it was transferred to a given solution and its capacitive response is recorded between −0.5 and 0.8 V. Then, the electrode was subjected to the reduction (potentiostatic, 10 s) at different potentials ($E_{red}$) between −0.8 and −1.6 V. After each reduction capacitance was measured between −0.5 and 0.8 V. Capacitance increment factors (CIF) are evaluated as the capacitance of rGO reduced at given $E_{red}$ and the capacitance of a starting, non-reduced GO. Knowing the mass of GO loaded on the electrode ($m$) specific capacitances were calculated as $C_{spec} = q/(2 \times m \times \Delta V)$ ($q$ is the charge within the cyclic voltammogram and $\Delta V$ is the width of potential window in which capacitance is measured, 1.3 V).

## 3. Results and discussion

The GO sample used in this work possesses high concentration of oxygen functional groups (C:O ratio close to 1) and a small concentration of heteroatoms (S and N). Due to high oxidation degree it is non conductive, as known for GO [19,23]. It undergoes irreversible electrochemical reduction, which is observed under both potentiodynamic and potentiostatic conditions. Reduction can be performed without conductive additive, as can be seen from the cyclic voltammograms of GO reduction in MCl and $MCl_2$ solution (M is alkali or alkaline earth metal) in Fig. 1 (top panel).

It is clearly seen that the GO reduction commences at lower potentials in $MCl_2$ solutions compared to MCl solutions and that it strongly depends on the cation present in the electrolyte. Reduction peak potential is the most negative for LiCl (−1.4 V) and shifts to more positive potentials for NaCl and KCl. For earth alkali metals reduction peak potential is at least 0.1 V shifted to more positive potentials. Previously we observed that the peak potential of the potentiodynamic reduction correlates with $E_{red}$ at which the highest CIF is observed after potentiostatic reduction when different GOs are used [19]. As can be seen in Fig. 1, such behaviour is preserved when reduction is performed in different electrolytes. When GO is reduced in MCl and MCl$_2$ solutions, the initial capacitance (ranging from 1.5 to 3.2 F g$^{-1}$, depending on the electrolyte) increases 20 to 70 times (at $E_{red}$ corresponding to maximum CIF).

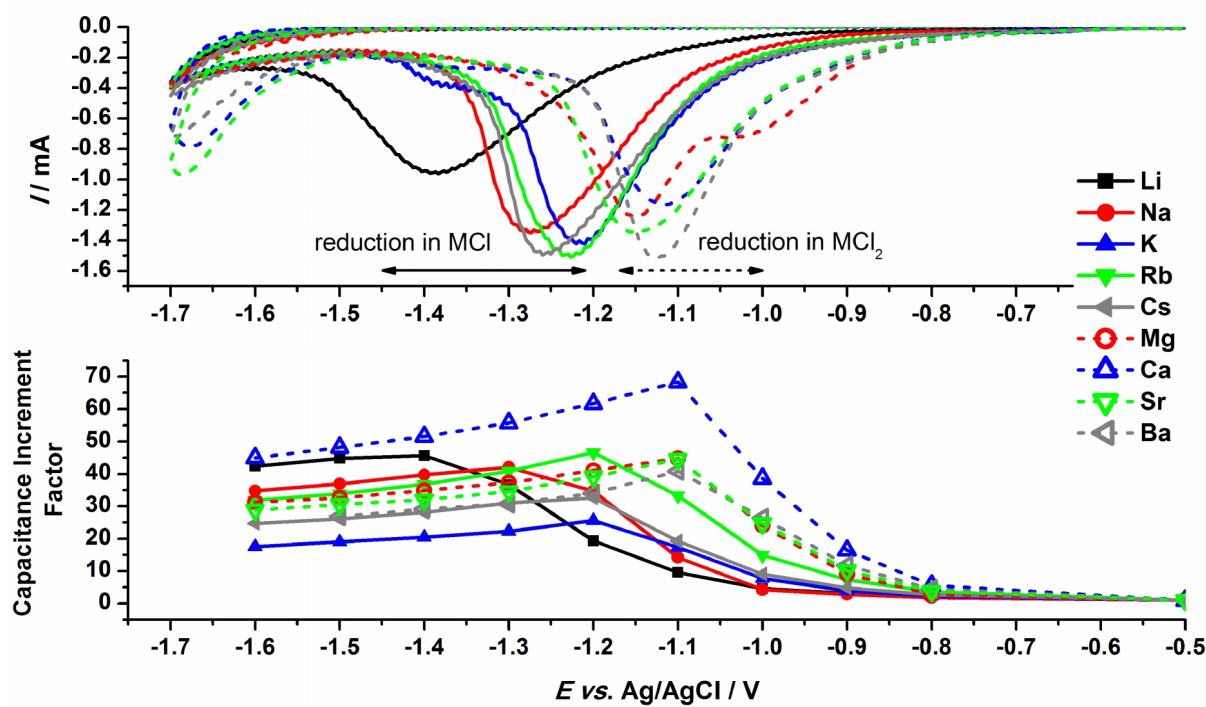

**Figure 1.** top panel: cyclic voltammograms of GO reduction in MCl and MCl$_2$ solutions (0.1 mol dm$^{-3}$) (full lines – alkali metals, dashed lines – alkaline earth metals); bottom panel: capacitance increment factors as the function of reduction potential. For the presented measurements no conductive additive is used.

Beside a clear dependence of the reduction process on the composition of the "inert" electrolyte and separation of the reduction peaks between alkali and alkaline earth metals, it is difficult to observe some distinct trend in Fig. 1. However, when materials for electrochemical capacitors are used it is a common practice to include some conductive additive to the electrode formulation, in order to improve current collection efficiency. In the second set of experiments we used this strategy and investigated GO reduction in MCl and MCl$_2$ solutions when conductive additive is included in the GO film (Fig. 2). Again, reduction in MCl solutions takes place at lower potentials compared to MCl$_2$ solution. When compared to the case where no conductive additive is included (Fig. 1), in all the cases reduction takes place at more positive potentials. We ascribe this effect to a higher conductivity of GO films in this series of experiments.

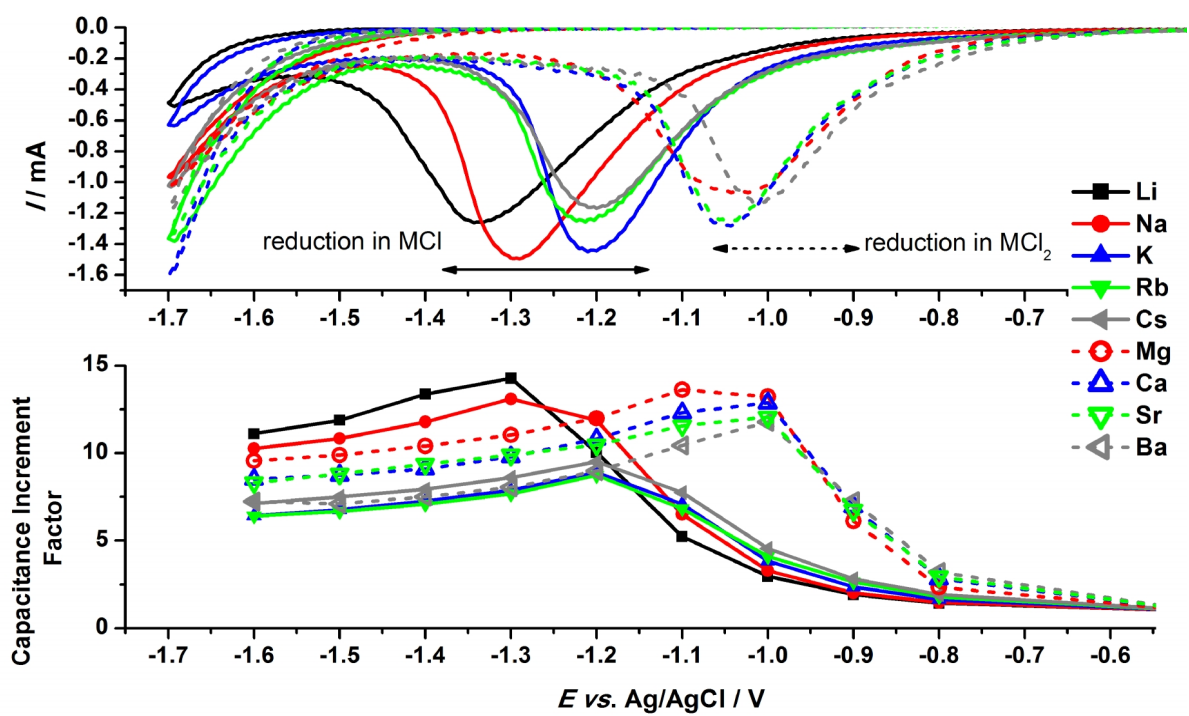

**Figure 2.** top panel: cyclic voltammograms of GO reduction in MCl and MCl$_2$ solutions (0.1 mol dm$^{-3}$) (full lines – alkali metals, dashed lines – alkaline earth metals); bottom panel: capacitance increment factors as the function of reduction potential. GO films contained Vulcan XC-72R as conductive additive.

Moreover, in Fig. 2 it is clearly seen that, in general, reduction peak potential shift to more positive potentials as the crystallographic radius of cation increases (ie, hydrated ion radius decreases [24]) in the series of alkali metals and alkaline earth metals. In the same manner CIFs decrease. Moreover, when compared to the results given in Fig. 1 (bottom panel) it can be observed that the CIFs are significantly lower when conductive additive is used. As previously explained [19] this is not a negative effect of the conductive carbon added to the GO layer, but the consequence of higher initial GO capacitances due to a higher conductivity of the GO/conductive component film. Namely, we measured the initial specific capacitances (i.e. before the reduction) of GO/Vulcan XC-72R films around 10 F g$^{-1}$.

In Fig. 3(a) we show the evolution of capacitive response of the GO film reduced in KCl solution at different potentials. Development of pseudocapacitance is seen as a wide hump observed in the potential window −0.5 to 0.3 V. Evaluated specific capacitances for different $E_{red}$ are shown in Figs.3(b) and (c).

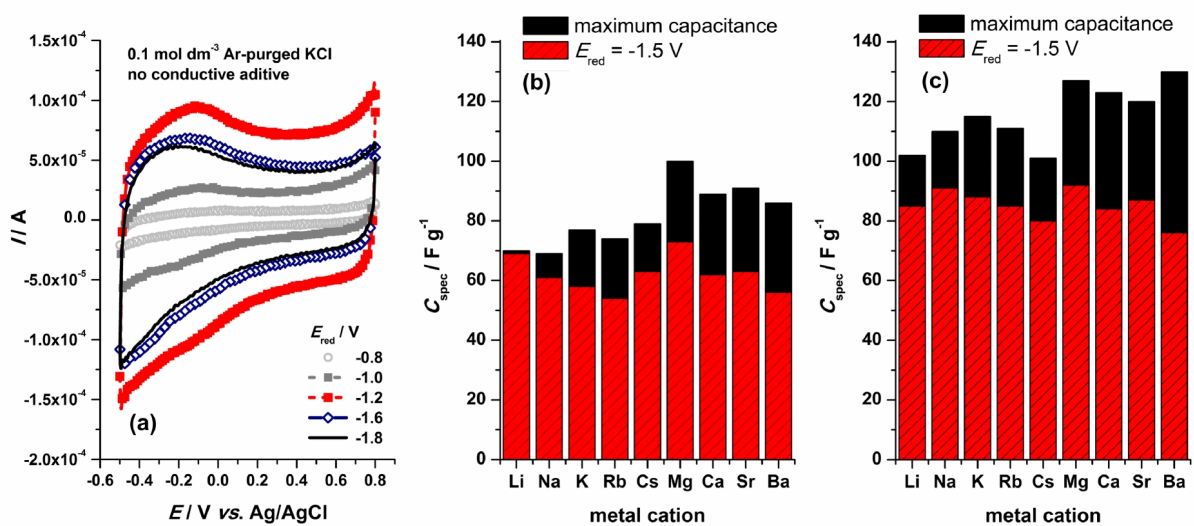

**Figure 3.** (a) evolution of the capacitive response of GO upon reduction in KCl solution at different reduction potentials ($E_{red}$; no conductive additive); (b) Maximum specific capacitances ($C_{spec}$) of GO (corresponding to capacitances measured at $E_{red}$ which corresponds to the maximum CIF) and $C_{spec}$ measured after reduction at −1.5 V in the absence of conductive additive; (c) the same as for (b) but with conductive additive in the GO film.

$C_{spec}$ obtained upon the reduction show dispersion between 65 and 130 F g$^{-1}$ (values corresponding to the maximum CIFs, Figs. 1 and 2). This demonstrates efficient capacitance tuning via electrochemical reduction. Capacitances measured after the reduction in MCl solutions are typically lower than the ones measured in $MCl_2$ solutions. As in all the cases maximum CIFs correspond to the peak potential of the potentiodynamic GO reduction, it is not surprising why in the alkali metal chlorides and alkaline earth metal chlorides series we observe similar values of capacitances. In all of the cases maximum capacitance is reached when the conductivity is balanced with the concentration of oxygen functional groups, which contribute capacitance through the pseudofaradaic processes. Here we observe that this state corresponds to roughly 60% of oxygen functional groups removed from the GO surface, irrespective of the electrolyte used (evaluated through the charge under the cyclic voltammograms of GO reduction). Variations of $C_{spec}$ are due to the specific interactions of $M^+$ and $M^{2+}$ ions with the remaining oxygen functional groups on the reduced GO surface. Also, $C_{spec}$ measured in the presence of conductive additive are higher than the ones measured upon the reduction of pure GO films. We explain this result by higher conductivity of reduced GO/Vulcan XC-72R films and more opened structure of the films containing Vulcan XC-72R. Namely, 50 nm globular particles of Vulcan XC-72R [25] prevent extensive stacking of GO layers and provide more access to the electrolyte during the potentiodynamic cycling.

Here we emphasize clear differences in the observed GO reduction peak potentials (Figs. 1 and 2, upper panels) in different electrolytes. Irrespective on the addition of the conductive component, the reduction takes place at lower potentials in earth alkali metal chlorides. Moreover, in the presence of Vulcan XC72-R reduction potential shifts to more positive potentials when going from LiCl to NaCl and KCl, while in RbCl and CsCl peak potentials are close to the one in KCl. In the case of the $MCl_2$ solutions the peak potentials are close for $MgCl_2$, $CaCl_2$ and $SrCl_2$, while it is more positive for $BaCl_2$. As the reduction is performed at relatively deep cathodic potentials, the explanation of this behaviour can be sought in the interactions of cations with the oxygen functional groups of GO. Theoretical considerations of the alkali metals-oxygen functional groups interactions [8] have shown the charge transferred from the metal atom localizes in vicinity of the oxygen functional group which interacts with alkali metal atom. Based on these results, we expect that doubly charged alkaline earth metal cations strongly localize negative charge brought to the GO layer and in this way activate the reduction of the oxygen functional groups and their removal from the GO surface more efficiently than singly charged alkali metal cations. This explains the reduction of GO at lower potentials when performed in $MCl_2$ solutions. Considering observed trends in the series of MCl and $MCl_2$, we expect that ions of the same charge (1+ or 2+) with larger hydratation spheres localize charge to lower extent, as they are located at greater distance from the GO surface. For this reason, reduction in LiCl commences at more negative potentials compared to the cases of NaCl and KCl. $Cs^+$ and $Rb^+$ have similar radii of hydrated ions like $K^+$ (around 230 pm [24]) and for this reason reduction peak potentials are close in these three cases (Fig. 2, upper panel).

## 4. Conclusion

We have shown that the electrochemical reduction of GO is highly sensitive to the type of cations present in the electrolyte. Reduction of GO takes place at lower potentials in alkali metal chloride solution, compared to the earth alkali metal chlorides. Upon reduction capacitance increases dramatically, but the $E_{red}$ of the maximum capacitance is dependent of composition of the electrolyte. Using electrochemical reduction maximum capacitances in the range 65 to 130 F $g^{-1}$ are obtained, which also depended on the presence of conductive component in the GO film. We propose that observed effect is due to specific interactions of cations with present oxygen which can activate removal of oxygen functional groups at different potentials.

**Acknowledgements**

I.A.P. acknowledges the support provided by Serbian Ministry of Education, Science and Technological Development (project III45014). Authors are grateful to Denis Sačer from Faculty of Chemical Engineering and Technology, University of Zagreb, for providing us with the graphene oxide sample.